\documentclass[aps,prb,twocolumn,showpacs,draft]{revtex4}

\usepackage{bm,theorem}

\newtheorem{post}{Postulate}

\renewcommand{\theequation}{\arabic{section}.\arabic{equation}}
\newcommand{\ab}[1]{\setcounter{equation}{0}\section{#1}}
\newcommand{\gl}[1]{(\ref{#1})}

\begin{document}

\title{Nonadiabatic extension of the Heisenberg model}

\author{Ekkehard Kr\"uger} 
\email[]{krueger@physix.mpi-stuttgart.mpg.de}
\affiliation{Max-Planck-Institut f\"ur Metallforschung, D-70506 Stuttgart,
  Germany} 
\date{\today}

\begin{abstract}
  The localized states within the Heisenberg model of magnetism should be
  represented by best localized Wannier functions forming a unitary
  transformation of the Bloch functions of the narrowest partly filled energy
  bands in the metals.  However, as a consequence of degeneracies between the
  energy bands near the Fermi level, in any metal these Wannier functions
  cannot be chosen symmetry-adapted to the complete paramagnetic group $M^P$.
  Therefore, it is proposed to use Wannier functions with the reduced
  symmetry of a magnetic subgroup $M$ of $M^P$ [case (a)] or spin dependent
  Wannier functions [case (b)]. The original Heisenberg model is
  reinterpreted in order to understand the pronounced symmetry of these
  Wannier functions. While the original model assumes that there is exactly
  one electron at each atom, the extended model postulates that in narrow
  bands there are as many as possible atoms occupied by exactly one electron.
  However, this state with the highest possible atomiclike character cannot
  be described within the adiabatic (or Born-Oppenheimer) approximation
  because it requires a more realistic description of the electronic motion.
  Within the (true) nonadiabatic system the electrons move on localized
  orbitals that are still symmetric on the average of time, {\em but not} at
  any moment. These nonadiabatic states have the same symmetry as the
  adiabatic states and determine the commutation properties of the
  nonadiabatic Hamiltonian $H^n$.  The nonadiabatic Heisenberg model is a
  purely group-theoretical model which interprets the commutation properties
  of $H^n$ that are explicitly given in this paper for the two important
  cases (a) and (b). There is evidence that the occurrence of these two types
  of Wannier functions in the band structure of a metal is connected with the
  occurrence of magnetism and superconductivity, respectively.
\end{abstract}
\pacs{Heisenberg model, 75.10.-b, 74.20.-z}

\maketitle

\ab{Introduction} 

In any application of the Heisenberg model of magnetism,\cite{hei} the
localized states of the electrons should be represented by Wannier functions
\begin{equation}
  \label{wf}
w_i(\vec r - \vec R - \vec\rho_i) = \frac{1}{\sqrt{N}}\sum^{BZ}_{\vec
  k}\sum_{q = 1}^{\mu}e^{-i\vec k (\vec R + \vec\rho_i)} g_{iq}(\vec k
  )\varphi_{\vec kq}(\vec r)   
\end{equation}
with the following properties:

(i) the $w_i(\vec r - \vec R - \vec\rho_i)$ are centered on the atomic
positions $\vec R + \vec\rho_i$;

(ii) the $w_i(\vec r - \vec R - \vec\rho_i)$ are gained by a {\em unitary}
  transformation from the Bloch functions $\varphi_{\vec kq}(\vec r)$ of
  the energy bands of interest; 
  
  (iii) the $w_i(\vec r - \vec R - \vec\rho_i)$ are symmetry-adapted to the
  space group $G$ of the considered metal; and
  
  (iv) the $w_i(\vec r - \vec R - \vec\rho_i)$ are as well localized as
  possible. 
  
The first sum in Eq.~\gl{wf} runs over the $N$ vectors $\vec k$ of the
first Brillouin zone (BZ), the second sum runs over the $\mu$ bands of
interest (with the band indices $q = 1$ to $\mu$), and $\vec R$ and
$\vec\rho_i$ denote the vectors of the Bravais lattice and the centers of
symmetry of the Wannier functions within the unit cell, respectively.

The transformation is unitary [point (ii)] if the coefficients $g_{iq}(\vec
k)$ in Eq.~\gl{wf} are the elements of a unitary matrix ${\bm g}(\vec k)$,
\begin{equation}
  \label{gk}
{\bm g}^{-1}(\vec k) = {\bm g}^{\dagger}(\vec k).  
\end{equation}
The Wannier functions are symmetry-adapted to $G$ [point (iii)] if
they satisfy the equation
\begin{equation}
  \label{swf}
w_i\big(\alpha^{-1}(\vec r - \vec R - \vec\rho_i)\big) = \sum_{j =
  1}^{\mu}D_{ji}(\alpha )w_j(\vec r - \vec R - \vec\rho_i) 
\end{equation}
for the elements $\alpha$ of the point group $G_0$ of $G$, where the matrices
$[D_{ji}(\alpha )]$ form a (reducible or irreducible) single-valued
representation $D_0$ of $G_0$, see Eq.~(1.8) of Ref.~\onlinecite{ew1}.
[Note that in Eq.~\gl{swf} on the right hand side there is $\rho_i$ and not
$\rho_j$.]

The Wannier functions are as well localizable as possible [point (iv)] if the
Bloch-like functions
\begin{equation}
  \label{eq:26}
\widetilde\varphi_{\vec ki}(\vec r) = \sum_{q = 1}^{\mu}g_{iq}(\vec k
  )\varphi_{\vec kq}(\vec r)     
\end{equation}
vary (for fixed $\vec r$) smoothly through the whole $\vec k$
space.\cite{ew1} 

It is one of the most important results of the group theory of Wannier
functions that Wannier functions complying with all the four conditions given
above exist only in {\em isolated} sets of $\mu$ energy bands which satisfy
the group-theoretical compatibility relations throughout the Brillouin
zone.\cite{dclI,dclII,dclIII,ew1,ew2} [This condition is necessary, but not
sufficient: in addition, there must exist unitary matrices ${\bm S}(\vec K)$
defined in Eq.~(4.16) of Ref.~\onlinecite{ew1} which satisfy the equations
(4.17) and (4.28) of Ref.~\onlinecite{ew1}. These matrices ${\bm S}(\vec K)$
determine the positions $\vec\rho_i$ of the Wannier functions.] In
an ``isolated'' set of energy bands, each band may be connected by
degeneracies to the other bands of this set, but must not be connected to
bands not belonging to this set.

The mentioned ``energy bands of interest'' are the partly filled energy bands
in the band structures of the considered metal. Often, it is only one roughly
half-filled band which interests. However, the energy bands in the
(paramagnetic) band structures of the metals are degenerate at several
points and lines of symmetry of the Brillouin zone. Because of these
degeneracies, is not possible to separate narrow isolated sets of energy
bands which satisfy the compatibility relations throughout the Brillouin
zone.  For this reason, Wannier functions with all the properties demanded
above {\em do not exist in the metals}.

Therefore, the localized states are often represented by ``approximated''
Wannier functions which no longer form an exactly unitary transformation of
the Bloch functions. These Wannier functions are constructed from slightly
modified energy bands in which some of the Bloch functions at points, lines,
and planes of symmetry are replaced by Bloch functions with a symmetry
appropriate for the construction of Wannier functions. Hence, these
approximated Wannier functions have lost all the information connected with
the symmetry of the removed Bloch functions and carry the wrong information
of the new Bloch functions.

The nonadiabatic Heisenberg model (NHM) as proposed in this and previous
papers\cite{es,ea,ef,em} extends the original Heisenberg model on the
basis of Wannier functions which form an {\em exactly} unitary transformation
of the Bloch functions of the bands of interest. Within this model, {\em it
  is not allowed} to replace any Bloch function in the calculated band
structure by functions with a new symmetry. Hence, the NHM takes into account
the {\em complete} information connected with the symmetry of the Bloch
functions in the band structure of the considered metal.

Clearly, the Wannier functions used within the NHM cannot not comply with all
the properties (i) -- (iv) given above. The development of the nonadiabatic
model was suggested by two observations:
 
1. An exactly unitary transformation of the Bloch functions of the partly
filled bands into best localized Wannier functions becomes possible in nearly
all the metals when the Wannier functions are allowed to have the {\em
  reduced symmetry} of a magnetic subgroup $M$ of the paramagnetic group [see
appendix~\ref{app:a}, case (a)] or when they are allowed to be {\em spin
  dependent} [see appendix~\ref{app:a}, case (b)].

2. There is a deep connection between the symmetry of these {\em exact}
Wannier functions in a metal and the physical properties of the electrons
at the Fermi level.\cite{es2,es,em,ef,ea}

The original Heisenberg model of magnetism is defined by the assumption that
there is exactly one electron on each atom of a metal.  The NHM replaces this
assumption by introducing three postulates which will be given in the
following Sec.~\ref{sec2}. These postulates combine in a new way the
Heisenberg model with the band model. The fundamental second postulate given
in Eq.~\gl{hcbn0} states that the probability to find exactly one electron on
an atom is as {\em large as possible} in narrow energy bands.

The second postulate of the NHM cannot be satisfied within the adiabatic (or
Born-Oppenheimer) approximation. In the framework of this approximation the
electrons move in {\em rigid} orbitals in the {\em average} potential of the
other electrons. The second postulate, however, requires a more realistic
description of the electronic motion.  In the true (nonadiabatic) system a
localized electron moves in a potential depending on which of the adjacent
localized states is occupied and on the present motion of the electrons in
these states. These modified orbitals
$$
|\vec T,m,\nu \rangle  
$$
are described by introducing a new quantum number $\nu$ which labels
different states of motion of the center of mass of the localized
states.\cite{bem2} [$\vec T$ denotes the positions of the atoms, see
Eq.~\gl{tr}.]

Nonadiabatic localized functions 
$$\langle \vec r,t,\vec q\,|\vec T, m, \nu\rangle$$
(as introduced in the
next section) which represent the nonadiabatic localized states are highly
complicated. Hence, it will be practically impossible to give these functions
explicitly. Fortunately, one important feature of these functions is known
exactly: they have the same symmetry as the {\em exact} Wannier functions of
the narrowest, roughly half-filled energy bands of the metal under
consideration.  Thus, any application of the NHM starts with a
group-theoretical examination of the symmetry of the best localized (spin
dependent) Wannier functions which is clearly determined by the symmetry of
the Bloch functions in the band structure of the given
metal.\cite{dclI,dclII,dclIII,ew1,ew2,ew3} The symmetry of these Wannier
functions is explicitly given for two important cases (a) and (b) in appendix
\ref{app:a}.

The NHM is a purely group-theoretical model. An explicit knowledge of
the nonadiabatic localized functions (going beyond of their symmetry) does
not provide new physical insight. Even in the nonadiabatic model, any
calculation of expectation values should be carried out within the adiabatic
approximation.

\ab{Nonadiabatic Heisenberg model}
\label{sec2}
\subsection{General}
Consider a set of $\mu$ energy bands in a metal with the paramagnetic space
group $G$, the paramagnetic group
\begin{equation}
\label{pmg}
M^P = G + KG  
\end{equation}
(with $K$ denoting the operator of time inversion), and $\mu$ atoms at the
positions
\begin{equation}
\vec T = \vec R + \vec\varrho_{i}
\label{tr}
\end{equation}
per unit cell, where $\vec R$ and $\vec\varrho_{i}$ ($i = 1$ to $\mu$) denote
the vectors of the Bravais lattice and the positions of the $i$th atom within
the unit cell, respectively. The energy bands of this set are assumed to
belong to the narrowest partly filled bands of this metal (while it is not
demanded that {\em all} the narrow, partly filled bands belong to it).

Assume that the symmetry of the Bloch functions of the considered set of
energy bands allows the construction of {\em either} Wannier
functions
\begin{equation}
  \label{uswf}
w_{\vec Tm}(\vec r, t) \equiv w_{i}(\vec r - \vec R - \vec\rho_i)u_m(t)
\end{equation}
symmetry-adapted to a magnetic subgroup $M$ of $M^P$ 
{\em or} spin dependent Wannier functions
\begin{equation}
  \label{eq:6}
w_{\vec Tm}(\vec r, t) \equiv w_{im}(\vec r - \vec R - \vec\rho_i, t)
\end{equation}
symmetry-adapted to $M^P$. The former functions are defined
appendix~\ref{app:a}, case (a), their symmetry is given in
Eqs.~\gl{pswf3} and \gl{kswf2}, the latter are defined in
appendix~\ref{app:a}, case (b), and their symmetry is given in Eqs.~\gl{ssdwf} and
\gl{ksdwf}. The functions $u_m(t)$ are Pauli's spin functions, see
Eq.~\gl{paulisf}, $t$ is the spin coordinate, and the (crystal) spin label $m
= \pm\frac{1}{2}$ distinguished between the two functions at the same
position $\vec T$. In either case, the Wannier functions form a unitary
transformation of the {\em exact} Bloch functions of the considered set of
$\mu$ energy bands, are situated on the atoms (with the positions $\vec T$),
and are as well localized as possible.

\subsection{The three postulates of the nonadiabatic Heisenberg model}
\label{post}

Let be
\begin{equation}
H = H_{HF} + H_{Cb}
\label{h}
\end{equation}
the electronic Hamiltonian in the considered set of energy bands with $H_{HF}$ and
\begin{eqnarray}
H_{Cb}& = &\sum_{\vec T, m}\langle\vec T_{1}, m_1; \vec T_{2}, m_2|H_{Cb}|
\vec T_{1}', m_1'; \vec T_{2}', m_2'\rangle\nonumber\\
&&\times c_{\vec T_{1}m_{1}}^{\dagger}
c_{\vec T_{2}m_{2}}^{\dagger}
c_{\vec T_{2}'m_{2}'}
c_{\vec T_{1}'m_{1}'}
\label{hcb}
\end{eqnarray}
representing the Hartree-Fock and Coulomb energy, respectively.  The fermion
operators $c_{\vec Tm}^{\dagger}$ and $c_{\vec Tm}$ create and annihilate
electrons with (crystal) spin $m$ in the localized states $|\vec T, m\rangle$
represented by the Wannier functions $w_{\vec Tm}(\vec r, t)$. Other
contributions to $H$ from the electrons not belonging to the considered set
of bands, are neglected even as are spin-orbit effects.

$H_{Cb}$ may be written as
\begin{equation}
H_{Cb} = H_{c} + H_{ex} + H_{z},
\label{cz}
\end{equation}
with the operator of Coulomb repulsion $H_{c}$ containing all the matrix
elements of $H_{Cb}$ with
\begin{equation}
\vec T_{1} = \vec T_{1}' ~\mbox{and}~ \vec T_{2} = \vec T_{2}',
\label{r12}
\end{equation}
the exchange operator $H_{ex}$  containing the matrix elements with
\begin{equation}
\vec T_{1} = \vec T_{2}'~ \mbox{and}~ \vec T_{2} = \vec T_{1}',
\label{r21}
\end{equation}
and $H_{z}$ comprising the remaining matrix elements, i.e., the matrix
elements with
\begin{equation}
\{\vec T_1,\vec T_2\} \neq \{\vec T_1',\vec T_2'\}.
\label{r12r21}
\end{equation}

The interaction $H_z$ is of great importance within the NHM. In order to
discuss the effect of $H_z$, consider the operator 
\begin{equation} 
H' = H_{HF} + H_{c} + H_{ex}
\label{h'}
\end{equation} 
obtained from the complete Hamiltonian $H$ by putting $H_z = 0$. 

As is well-known, the Coulomb repulsion of two electrons occupying localized
states at the same atom is larger than the Coulomb repulsion of two electrons
at different atoms.\cite{mott} The electronic motion in the ground state
$|G\,'\rangle$ of $H'$ has an ``atomiclike'' character when the Coulomb
repulsion between the localized states determines the electronic motion in
$|G\,'\rangle$ to such an extend that the probability to find two electrons
(with different spin directions) on the same atom is markedly smaller than in
case of a purely bandlike motion. In this context, I speak of a ``purely
bandlike'' motion when the probability to find an electron in the localized
state $|\vec Tm\rangle$ is independent of whether or not the other state
$|\vec T, -m\rangle$ is occupied. In this case, the ground state consists of
configurations with nearly random occupation.

The atomic- or bandlike character depends on the mean time of stay $\tau
\approx \hbar/\Delta$ of the electrons at the atoms and, hence, on the
bandwidth $\Delta$. For $\Delta \rightarrow 0$ we have $\tau \rightarrow
\infty$; the metal becomes a Mott insulator representing a perfectly
atomiclike state. For $\Delta \rightarrow \infty$, on the other hand, we have
$\tau \rightarrow 0$ and, hence, a purely bandlike character of the
electrons. Thus, as is well-known, the electrons in partly filled energy
bands tend to a more atomiclike behavior with decreasing bandwidth and to a
more bandlike behavior with increasing bandwidth.\cite{hubbard}

Now assume the considered energy bands to be sufficiently narrow that the
ground state $|G\,'\rangle$ of $H'$ clearly has atomiclike character.  The
matrix elements of $H_{z}$ satisfy neither Eq.~\gl{r12} nor \gl{r21}.  Thus,
the interaction $H_{z}$ annihilates two electrons in localized
states at the positions $\vec T_{1}'$ and $\vec T_{2}'$ and creates at least
one of them at the {\em new} positions $\vec T_{1}$ or $\vec T_{2}$. Hence,
unlike $H_{c}$ or $H_{ex}$, the operator $H_{z}$ generates transitions
between adjacent localized states which lead to configurations with a more
random occupation. Consequently, the probability to find two electrons at the
same position $\vec T$ will be larger in the ground state $|G\rangle$ of the
complete Hamiltonian
$$H = H' + H_z$$
[given in Eq.~\gl{h}] than in the ground state
$|G\,'\rangle$ of $H'$. Therefore, the total Coulomb repulsion energy in
$|G\rangle$ is larger than in $|G\,'\rangle$ and we may assume that in
sufficiently narrow bands the ground state energy $E$ of $H$ is greater than
the ground state energy $E'$ of $H'$.

Still ``sufficiently'' narrow means that the Coulomb repulsion between the
localized states determines the electronic motion in $|G\,'\rangle$.  It is
difficult to decide how narrow such sufficiently narrow bands should be.
However, we know that there is strong theoretical and experimental evidence
that, e.g., the $d$ electrons of the transition metals exhibit behavior of
both the band and the Heisenberg model.\cite{hubbard} Therefore, we may
suppose that $E > E'$ is valid in the {\em narrowest} bands of the
metals. This supposition leads to the first postulate of the NHM.

\begin{post}
   In the narrowest, partly filled energy bands of
  the metals the transitions generated by $H_z$ are energetically
  unfavorable, i.e., we have 
\begin{equation}
\langle G|H|G\rangle > \langle G\,'|H'|G\,'\rangle,
\label{hgh'}
\end{equation}
\nopagebreak
where $|G\rangle$ and $|G\,'\rangle$ denote the {\bf exact} ground
states of $H$ and $H'$, respectively. 
\end{post}

The particular form of the matrix elements of $H_{z}$ shows that it
represents a short-ranged interaction which crucially depends on the exact
form of the localized functions. This fact suggests that only small changes
of the localized electronic orbitals are required to prevent the transitions
generated by $H_{z}$. However, such modified orbitals do not exist within the
adiabatic approximation because such modifications yield localized charge
distributions which are not symmetric at any moment. As a consequence, the
nuclei become accelerated in varying directions. Hence, we replace the
(adiabatic) localized states (represented by the Wannier functions) by more
realistic nonadiabatic localized states
\begin{equation}
|\vec T,m,\nu \rangle 
\label{nls}
\end{equation}
which take into account the motion of the nuclei. The new quantum number
$\nu$ labels different states of motion of the center of mass of the nucleus
and the electron occupying the state $|\vec T, m, \nu \rangle$.\cite{bem2}

The nonadiabatic Hamiltonian $H^{n}$ may be written as
\begin{equation}
H^{n} = H_{HF} + H_{Cb}^{n}
\label{hn}
\end{equation}
where the Coulomb interaction now has the form
\begin{eqnarray}
H_{Cb}^{n} &=&
\sum_{\vec T, m}\langle\vec T_{1}, m_{1}, n; 
\vec T_{2}, m_{2}, n|H_{Cb}|
\vec T_{1}', m_{1}', n; \vec T_{2}', m_{2}', n \rangle\nonumber\\
&&\times c_{\vec T_{1}m_{1}}^{n\dagger}
c_{\vec T_{2}m_{2}}^{n\dagger}
c_{\vec T_{2}'m_{2}'}^{n}
c_{\vec T_{1}'m_{1}'}^{n}.
\label{hcbn}
\end{eqnarray}
The new fermion operators $c_{\vec Tm}^{n\dagger}$ and $c_{\vec Tm}^{n}$
create and annihilate electrons with crystal spin $m$ [see
appendix~\ref{app:a}, case (b)] in the nonadiabatic localized states $|\vec
T, m, n \rangle$. The matrix elements of $H_{Cb}^{n}$ are integrals
\begin{eqnarray}
\lefteqn{\langle\vec T_{1}, m_{1}, n; 
\vec T_{2}, m_{2}, n|H_{Cb}|
\vec T_{1}', m_{1}', n; \vec T_{2}', m_{2}', n \rangle}\nonumber\\*
&=&\frac{e^2}{2}\sum_{tt'}\int\langle\vec T_{1}, m_{1}, n| 
\vec r,t,\vec q~\rangle\langle\vec T_{2}, m_{2}, n| 
\vec r\,',t',\vec q\,'~\rangle\nonumber\\*
&&\times\frac{1}{|\vec r - \vec r\,'|}\langle\vec r,t,\vec q~|\vec T'_{1}, m'_{1},
n\rangle \langle\vec r\,',t',\vec q\,'|\vec T'_{2}, m'_{2},
n\rangle\nonumber\\*
&&\times d\vec rd\vec r\,'d\vec qd\vec q\,'
\end{eqnarray}
over nonadiabatic localized functions of the form  
\begin{equation}
\langle \vec r,t,\vec q~|\vec T, m, n \rangle,
\label{nalsd}
\end{equation} 
where $\nu = n$ labels the nonadiabatic states which satisfy the following
Eq. \gl{hcbn0}, and the new coordinate $\vec q$ stands for that
part of the motion of the center of mass of the localized
state $|\vec T, m, n \rangle$ which nonadiabatically follows the motion of
the electron occupying this state. We may imagine that $\vec q$ denotes 
the {\em acceleration} of the nucleus (together with the core electrons). 

Within the nonadiabatic localized states $|\vec T, m, \nu \rangle$ the
electrons possess considerably more room to move than within the adiabatic
states.  While in the adiabatic approximation the symmetry operators $P(a)$
act on $\vec r$ and $t$ alone, in the nonadiabatic system these operators act
on $\vec r$, $t$, {\em and} the acceleration $\vec q$ of the nuclei, see
Eq.~\gl{effectp}. The nonadiabatic localized functions have no definite
transformation properties under space group operations acting {\em only} on
$\vec r$ and $t$. Hence, the symmetry of the adiabatic and nonadiabatic
localized states [given by Eq.~\gl{eq:0} for the fermion operators] may be
interpreted as follows.

Within the adiabatic system the electrons move on orbitals being symmetric
with respect to the lattice {\em at any moment}.  Within the nonadiabatic
system, on the other hand, the orbitals are still symmetric on the average of
time, {\em but not} at any moment.  This statement is independent of the
absolute value $|\vec q\,|$ of the acceleration of the nuclei, i.e., it is
independent of whether or not the mass of the electrons is markedly smaller
than the mass of the nuclei.

Thus, the introduction of the new quantum number $\nu$ allows the electrons
to move in a potential depending on which of the adjacent localized states
are occupied and on the present positions of these electrons. Hence, within
the nonadiabatic system the electrons should be able to avoid the transitions
generated by $H_{z}$ by an appropriately modified motion, if these
transitions are energetically unfavorable, i.e., if the relation \gl{hgh'} is
true. Thus, as a consequence of relation \gl{hgh'}, all the matrix elements
of $H^n_{Cb}$ which neither satisfy Eq. \gl{r12} nor Eq. \gl{r21} should
vanish.

For this reason, we suppose that the transitions generated by $H_z$ are
artifacts of the adiabatic approximation and do not happen in the (true)
nonadiabatic system if relation \gl{hgh'} is satisfied. This supposition
leads to the second postulate of the NHM.

\begin{post}
  If relation \gl{hgh'} is true, the Coulomb interaction $H^n_{Cb}$ does not
  generate transitions between adjacent localized states, i.e.,
\begin{equation}
\langle\vec T_{1}, m_{1}, n; 
\vec T_{2}, m_{2}, n|H_{Cb}|
\vec T_{1}', m_{1}', n; \vec T_{2}', m_{2}', n \rangle
= 0
\label{hcbn0}
\end{equation}
for 
$$
\{\vec T_1,\vec T_2\} \neq \{\vec T_1',\vec T_2'\}
$$
and for special nonadiabatic localized functions
$$
\langle \vec r,t,\vec q~|\vec T, m, n \rangle
$$
labeled by $\nu = n$.
\end{post}

At the transition from the adiabatic to the nonadiabatic system, 
the total energy of the electron system decreases by
\begin{equation}
\Delta E = \langle G|H|G\rangle - \langle G\,'|H'|G\,'\rangle
\label{de}
\end{equation}
if we neglect the energy of the nonadiabatic motion of the nuclei and the
energy change caused by the slight modification of the electronic orbitals
within the nonadiabatic states.

As a consequence of Eq. \gl{hcbn0}, the commutation properties of the
operator $H^n_{Cb}$ depend on the symmetry of the nonadiabatic localized
states. Since only small modifications of the adiabatic electronic orbitals
are required to prevent the transitions generated by $H_{z}$, we can assume
that the nonadiabatic Hamiltonian $H^n$ has the same commutation properties
as the adiabatic Hamiltonian $H'$ given in Eq. \gl{h'}. This is the third
(and last) postulate of the NHM.

\begin{post}
  If relation \gl{hgh'} is true, the nonadiabatic Hamiltonian $H^n$ has the
  same commutation properties as the adiabatic Hamiltonian $H'$, i.e.,
\begin{equation}
[H',P]
\left\{
\begin{array}{c}
= \\
\neq
\end{array}
\right\}
0
\quad\Rightarrow\quad
[H^n,P]
\left\{ 
\begin{array}{c}
= \\
\neq
\end{array}
\right\}
0,
\label{equalcom}
\end{equation} 
where $P$ stands for any symmetry operator.
\end{post}
As a consequence, the
nonadiabatic localized functions have the same symmetry as the (adiabatic)
Wannier functions $w_{\vec Tm}(\vec r, t)$. 

\ab{Symmetry of the operator $H'$}
\label{sec:h'}

According to its definition, the operator $H'$ arises from the complete
adiabatic Hamiltonian $H$ in Eq.~\gl{h} by putting 
\begin{equation} 
H_z = 0.
\label{hz0}
\end{equation}
This equation does not state that $H_z$ is {\em neglected}, but that
$H_z$ is {\em put equal to zero}. By this step, the commutation properties of
the operator $H'$ depend on the symmetry of the Wannier functions, whereas
the commutation properties of the complete adiabatic Hamiltonian $H$ are
independent of the symmetry of the used basis functions.  The {\em
  nonadiabatic} matrix elements of $H_z$, however, vanish within the NHM, see
Eq.~\gl{hcbn0}.

\subsection*{Case (a): The Wannier functions are symmetry-adapted to a
  magnetic group}

If the Wannier functions are symmetry-adapted only to a magnetic subgroup $M$
of the paramagnetic group $M^P$, the symmetry of the operator $H'$ is given by
\begin{equation}
  \label{eq:7}
  [H',P(a)] = 0\qquad \mbox{for } a \in M
\end{equation}
and
\begin{equation}
  \label{eq:8}
  [H',P(a)] \neq 0\qquad \mbox{for } a \in (M^P - M),
\end{equation}
where
$$
M^P = G + KG
$$
stands for the paramagnetic group (and $G$ is the space group). The symmetry
operators $P(a)$ are given in Eq.~\gl{pa} and $K$ denotes the operator of
time inversion. Especially, in this case (a) $H'$ does not commute with $K$, 
\begin{equation}
  \label{eq:9}
  [H', K] \neq 0,
\end{equation}
since $K \in (M^P - M)$.

The first equation \gl{eq:7} is valid since the
complete Hamiltonian $H$ commutes with $P(a)$ and also the operator
$P(a)H'P^{-1}(a)$ complies with Eqs.~\gl{r12} and \gl{r21} if $a \in M$ since
the fermion operators $P(a)c^{\dagger}_{\vec Tm}P^{-1}(a)$ and $P(a)c_{\vec
  Tm}P^{-1}(a)$ are for all the $a \in M$ linear combinations of operators
$c^{\dagger}_{\vec T'm'}$ and $c_{\vec T'm'}$, respectively, labeled by the
{\em same} position $\vec T'$, see the equations \gl{eq:0} and \gl{eq:1}.

Within the NHM it is important that $H'$ does not commute with $P(a)$ for $a
\in (M^P - M)$. In the case (a) considered in this section, the Wannier
functions cannot be chosen in such a way that they are symmetry-adapted to a
group $\widehat M$ containing the operation $a$ as well as all the elements
of $M$. Consequently, for $a \in (M^P - M)$, the fermion operators
$P(a)c^{\dagger}_{\vec Tm}P^{-1}(a)$ do not comply with Eq.~\gl{eq:0} or
\gl{eq:1}, but are linear combinations
\begin{equation}
  \label{eq:27}
P(a)c^{\dagger}_{\vec Tm}P^{-1}(a) = \sum_{\vec T'm'}
d_{\vec T'm',\vec Tm}(\alpha )
c^{\dagger}_{\vec T'm'}
\end{equation}
of at least two operators $c^{\dagger}_{\vec T'm'}$ with {\em different}
labels $\vec T_1'$ and $\vec T_2'$.  We show that therefore the
operator $P(a)H'P^{-1}(a)$ has matrix elements violating Eq.~\gl{r12} or
\gl{r21}.

Consider a fermion operator
combination belonging to the Coulomb interaction of $H'$, say
\begin{equation}
  \label{eq:28}
O =
c_{\vec T_{1}}^{\dagger}
c_{\vec T_{2}}^{\dagger}
c_{\vec T_{2}}
c_{\vec T_{1}},
\end{equation}
and assume for a special $a \in (M^P - M)$
the sum in Eq.~\gl{eq:27} to consist of two summands,
\begin{equation}
  \label{eq:29}
Pc_{\vec T}^{\dagger}P^{-1} = a\cdot c_{\vec
  U}^{\dagger} + b\cdot c_{\vec V}^{\dagger},  
\end{equation}
labeled by the different positions $\vec U$ and $\vec V$. In Eq.~\gl{eq:29}
we use the abbreviation $P\equiv
P(a)$ and drop the index $m$ since it does not matter here. With
Eq.~\gl{eq:29} we obtain
\begin{eqnarray}
  \label{eq:30}
POP^{-1}\nonumber &=&
Pc_{\vec T_{1}}^{\dagger}
P^{-1}Pc_{\vec T_{2}}^{\dagger}
P^{-1}Pc_{\vec T_{2}}
P^{-1}Pc_{\vec T_{1}}
P^{-1}
\nonumber\\
&=&
(a c_{\vec U_1}^{\dagger} + b c_{\vec V_1}^{\dagger})
(a c_{\vec U_2}^{\dagger} + b c_{\vec V_2}^{\dagger})
\nonumber\\
&&\times
(a^* c_{\vec U_2} + b^* c_{\vec V_2})
(a^* c_{\vec U_1} + b^* c_{\vec V_1})
\end{eqnarray}
where $a \neq 0$ and $b \neq 0$.
For instance, the operator product
$$
c_{\vec U_{1}}^{\dagger}
c_{\vec V_{2}}^{\dagger}
c_{\vec U_{2}}
c_{\vec U_{1}}
$$
belonging to $POP^{-1}$ and, hence, to $PH'P^{-1}$, violates Eq.~\gl{r12}
since $\vec V_2 \neq \vec U_2$. Consequently, $H'$ does not commute with $P$
as expressed by Eq.~\gl{eq:8}. In the same way, $H'$ does not commute with
$P$ when there are more than two summands on the right hand side of
Eq.~\gl{eq:29}.

\subsection*{Case (b): The Wannier functions are spin dependent and
  symmetry-adapted to the paramagnetic group}

If we consider spin dependent Wannier functions symmetry-adapted to the
paramagnetic group $M^P$, then we have
\begin{equation}
  \label{eq:10}
  [H',P(a)] = 0\qquad \mbox{for } a \in M^P  
\end{equation}
and, especially,
\begin{equation}
  \label{eq:11}
  [H',K] = 0.
\end{equation}
However, in this case (b), the operator $H'$ has matrix elements with 
\begin{equation}
  \label{eq:25}
  m_1 + m_2 \neq m_1' + m_2'
\end{equation}
because the coefficients $f_{sm}(q,\vec k)$ in Eq.~\gl{sdbf} cannot be chosen
independent of $\vec k$, see appendix \ref{app:a}, case (b).  Therefore, $H'$
does not conserve the crystal spin and, hence\cite{es3}, does not commute
with the operators $M(\alpha )$ of the crystal spin defined in
Eq.~\gl{eq:19},
\begin{equation}
  \label{eq:12}
  [H',M(\alpha ) ] \neq 0 
\end{equation}
for at least one $\alpha \in G_M$.\cite{bem3}

\ab{Symmetry of the nonadiabatic Hamiltonian $H^{\lowercase{n}}$}
\label{sec:hn}

\subsection{Magnetic and paramagnetic group}
\label{sec:shn}

The nonadiabatic Hamiltonian $H^n$ has the same commutation properties as the
adiabatic operator $H'$, see Eq.~\gl{equalcom}. However, the symmetry
operators $P(a)$ now act on $\vec r, t$, and on the new
coordinate $\vec q$ of the nonadiabatic localized functions, see
Eq.~\gl{effectp}. 

Hence, we have
\begin{equation}
  \label{eq:7n}
  [H^n,P(a)] = 0\qquad \mbox{for } a \in M,
\end{equation}
\begin{equation}
  \label{eq:8n}
  [H^n,P(a)] \neq 0\qquad \mbox{for } a \in (M^P - M),
\end{equation}
and, especially, 
\begin{equation}
  \label{eq:9n}
  [H^n, K] \neq 0
\end{equation}
in the case (a) of the preceding section~\ref{sec:h'}, and
\begin{equation}
  \label{eq:10n}
  [H^n,P(a)] = 0\qquad \mbox{for } a \in M^P,  
\end{equation}
especially,
\begin{equation}
  \label{eq:11n}
  [H^n,K] = 0
\end{equation}
in the case (b) of the preceding section~\ref{sec:h'}.

\subsection{Crystal spin}
\label{sec:cs}

The nonadiabatic fermion operators in Eq.~\gl{hn} are no longer labeled by
the spin quantum number $s$. Hence, within the nonadiabatic system, the exact
Fermi excitations are no longer purely electronic states but
localized states of well-defined symmetry which are occupied by electrons
{\em carrying with them some nonadiabatic motion of the nuclei}. 

Let be $S(\alpha )$ with
\begin{equation}
  \label{eq:17}
S(\alpha )u_{s}(t) \equiv u_s(\alpha^{-1}t) = \sum_{s'} d_{s's}(\alpha
)u_{s'}(t)\quad\mbox{for }\alpha\in O(3)
\end{equation}
the operators turning the electron spin, where the functions
\begin{equation}
u_{s}(t) = \delta_{st}
\label{paulisf}
\end{equation}
are Pauli's spin functions with the spin quantum number $s = \pm \frac{1}{2}$
and the spin coordinate $t = \pm \frac{1}{2}$, and the matrices
$[d_{s's}(\alpha )]$ are the representatives of the two-dimen\-si\-onal
double-valued representation $D_{1/2}$ of the three-dimen\-si\-onal rotation
group $O(3)$.

The adiabatic Hamiltonian $H$ given in Eq.~\gl{h} commutes with the operators
$S(\alpha )$,
\begin{equation}
  \label{eq:23}
[H, S(\alpha )] = 0 \quad\mbox{for }\alpha \in O(3).  
\end{equation}
This equation expresses the conservation law of the spin angular momentum
within the adiabatic system.

The nonadiabatic Hamiltonian $H^n$, on the other hand, does not commute with
the operators $S(\alpha)$ (for $\alpha \neq E$) since the nonadiabatic
fermion operators are no longer labeled by the spin quantum number $s$.
Hence, as a consequence of the (small) shift of the Fermi character at the
transition from the adiabatic to the nonadiabatic system, the electron spin
angular momentum is no longer a conserved quantity. Now there exists an
interaction between the electron spins and the nonadiabatic motion of the
nuclei.

However, even in the nonadiabatic system there should exist a conserved
quantity related to the conservation law of angular momentum. 
Thus, the equation \gl{eq:23} should be replaced by an analogous equation
\begin{equation}
  \label{eq:24}
[{\cal H}^n, M(\alpha )] = 0\quad\mbox{for }\alpha \in G_M   
\end{equation}
in the nonadiabatic system, where ${\cal H}^n$ stands for the complete
nonadiabatic Hamiltonian. The group $G_M$ and the operators $M(\alpha )$ are
defined in Eqs.~\gl{eq:20} and \gl{eq:19}. They act on the quantum number $m$
of the nonadiabatic localized states $|\vec T, m, n\rangle$ in the same
manner as the operators $S(\alpha )$ act on the spin quantum number $s$ of
Pauli's spin functions $u_s(t)$, cf. Eq.~\gl{eq:22}. Therefore, these
operators may be called the symmetry operators of the ``crystal spin'' and
$m$ may be called the quantum number of the crystal spin. This is in analogy
to the wave vector $\vec k$ of the Bloch functions which is sometimes
referred to as ``crystal momentum'' in order to distinguish it from the
momentum $\vec p$.

In the case (b), i.e., if we consider spin dependent Wannier functions
symmetry-adapted to the paramagnetic group $M^P$, the adiabatic operator $H'$
does not commute with all the operators $M(\alpha )$, see Eq.~\gl{eq:12}.
Hence, also the nonadiabatic Hamiltonian $H^n$ as defined in Eq.~\gl{hn} does
not conserve the crystal spin,
\begin{equation}
  \label{eq:12n}
  [H^n,M(\alpha ) ] \neq 0 
\end{equation} 
for at least one $\alpha \in G_M$.\cite{bem3} It is one of the most
interesting problems of the NHM to interpret this equation, see
Sec.~\ref{outlook}, case (b).

\ab{Discussion}

\subsection{Crystal electrons}

The NHM has been developed in order to interpret the symmetry and spin
dependence of the Wannier functions in metals. These Wannier functions form
an {\em exactly} unitary transformation of the Bloch functions of a set of
partly filled energy bands in the band structure of the metal of interest.

In the framework of the adiabatic approximation, Wannier functions form
nothing but a unitary basis within the considered bands. Hence, the
commutation properties of the adiabatic Hamiltonian $H$ are independent of
the symmetry of the Wannier functions, since the symmetry of any adiabatic
Hamiltonian is independent of the symmetry of the used basis functions. The
symmetry and localization of the Wannier functions simplifies the calculation
of the matrix elements of $H$, but has no further physical meaning.

The nonadiabatic Hamiltonian $H^n$, on the other hand, has an important
feature which distinguishes it from any adiabatic Hamiltonian $H$: the
commutation properties and the spin dependence of $H^n$ depend on the
symmetry and spin dependence of the nonadiabatic localized functions. This is
because the nonadiabatic localized functions have a physical meaning going
beyond the meaning of pure basis functions: they represent states that are
really {\em occupied} by the electrons in the way described by
Mott\cite{mott} and Hubbard\cite{hubbard}: the electrons occupy the
nonadiabatic localized states as long as possible and perform their band
motion by hopping from one atom to another. Such a band motion is generally
referred to as atomiclike motion.

Within the NHM we may extend this picture of the atomiclike electron. Here
the whole localized state $|\vec T ,m,n\rangle$ behaves like a moving {\em
  particle}, say ``crystal electron'', with the local coordinate $\vec T$ and
the crystal spin $m$. The spatial extend of the crystal electron is
determined by the charge distribution of the localized state and the crystal
spin is a {\em conserved quantity}.

First consider the picture of the crystal electron within the adiabatic
approximation.  Both operators $H_c$ and $H_{ex}$ [given in Eq.~\gl{cz}]
represent interactions {\em between} crystal electrons and, hence, are in
accordance with this picture. The interaction $H_z$, on the other hand,
contradicts the picture of a moving crystal electron because it {\em
  destroys} these new particles.  Since $H_z$ is a short-ranged interaction,
we may say that within the adiabatic system ``the crystal electrons become
destroyed at the slightest touch''.

Within the NHM, on the other hand, Eq.~\gl{hcbn0} is valid. The crystal
electrons are {\em stable} in the nonadiabatic system because the Coulomb
interaction does not generate transitions between adjacent localized states.
We may interpret Eq.~\gl{hcbn0} by stating that ``the crystal electrons
become {\em slightly deformed} but not destroyed at a touch''. The
crystal electrons now have a certain {\em elasticity} protecting them from
being destroyed at any collision. In this context, the stability of crystal
electrons increases with decreasing band width.

\subsection{Outlook}
\label{outlook}

The symmetry of the nonadiabatic Hamiltonian $H^n$ is given in
Sec.~\ref{sec:hn} for two interesting cases (a) and (b) which shall be
considered separately.

\subsection*{Case (a): The Wannier functions are symmetry-adapted to a
  magnetic group}

Sets of narrow, roughly half-filled energy bands with Wannier functions
symmetry-adapted to a magnetic group $M$ as defined in the appendix
\ref{app:a}, case (a), have already been identified in the paramagnetic band
structures of iron\cite{ef} and chromium\cite{ea}. In both metals, this set
consists of one ``magnetic'' band.  In iron the related magnetic group
\begin{equation}
  \label{eq:13}
  M = I4/mm'm' = C^5_{4h} + K\{C_{2x}|\vec 0\}C^5_{4h}
\end{equation}
is the group of the ferromagnetic state and in chromium
\begin{equation}
  \label{eq:15}
  M = P_I4/mnc = D^6_{4h} + K\{E|\tau\}D^6_{4h}
\end{equation}
is the group of the commensurate spin-density-wave state.

In the case (a) considered in this section, the symmetry of the nonadiabatic
Hamiltonian $H^n$ is given by the Eqs.~\gl{eq:7n}, \gl{eq:8n}, and
\gl{eq:9n}. Equation~\gl{eq:9n} shows that $H^n$ does not commute with the
operator $K$ of time inversion. Therefore, $H^n$ cannot
have a paramagnetic or superconducting ground state, since both states are
invariant with respect to the time inversion.

Thus, the electrons of the magnetic band may gain the energy $\Delta E$
[given in Eq.~\gl{de}] only if the electron spins form a structure with the
magnetic group $M$. This fact may be interpreted as follows.\cite{ea,ef}

The electrons of the magnetic band {\em activate} a spin dependent exchange
mechanism producing a spin structure with the space group $M$. This is
possible since, first, the electrons can modify their orbitals in the
nonadiabatic localized states and, secondly, exchange integrals depend very
sensitively on the exact form of the electronic orbitals. Hence, the
electrons of the magnetic band modify their orbitals in such a way that the
exchange energy $E_{ex}$ is maximum for a spin structure with the group $M$. 

The condensation energy $E_f$, i.e., the energy difference between the
paramagnetic and the magnetic state, is no longer given by the exchange
energy $E_{ex}$ alone, but by
\begin{equation}
  \label{eq:16}
E_f = \Delta E +
E_{ex}.  
\end{equation}
Hence, $E_f$ may be positive even if $E_{ex}$ is negative.
  
\subsection*{Case (b): The Wannier functions are spin dependent and
  symmetry-adapted to the paramagnetic group}

Sets of narrow, roughly half-filled energy bands with spin dependent Wannier
functions symmetry-adapted to the paramagnetic group $M^P$ as defined in the
appendix \ref{app:a}, case (b), have already been identified in the band
structures of a great number of superconductors.\cite{es2,es,em} It is
remarkable that such ``superconducting'' bands cannot be found in those
metals (such as Li, Na, K, Rb, Cs, Ca Cu, Ag, and Au) which do not become
superconducting.\cite{es2}

In the case (b) considered in this section, the symmetry of the nonadiabatic
Hamiltonian $H^n$ is given by the Eqs.~\gl{eq:10n}, \gl{eq:11n}, and
\gl{eq:12n}. Eqs.~\gl{eq:10n} and \gl{eq:11n} show that the ground state of
$H^n$ has the correct symmetry of the paramagnetic group $M^P$. Especially,
$H^n$ commutes with the operator $K$ of time inversion. Therefore, $H^n$ may
have a paramagnetic or a superconducting ground state, but has not a magnetic
ground state.

Equation \gl{eq:12n} shows that $H^n$ does not conserve the crystal spin
angular momentum. Hence, the electrons of the considered bands may gain the
energy $\Delta E$ [given in Eq.~\gl{de}] only if they couple to other
excitations in such a way that the conservation of the crystal spin angular
momentum is satisfied in the nonadiabatic system. This fact may be
interpreted as follows.\cite{es,en,ehtc}

In isotropic materials, the electrons of the considered bands couple to the
phonons. This is possible since, first, the symmetry of localized acoustic
phonons shows that they are able to carry crystal spin angular momentum,
secondly, the electron spins are coupled to the phonons via the nonadiabatic
motion of the nuclei, and, thirdly, the resulting nonadiabatic Hamiltonian
complies with the conservation law of crystal spin angular momentum.

In anisotropic materials (consisting of one- or two-dimensional sublattices),
phonons are not able to transport crystal spin angular momenta through the
crystal.  Here the electrons of the considered bands are forced to couple to
energetically higher-lying boson excitations.

At zero temperature, this spin-boson interaction {\em constrains} the
electrons of the considered bands in a new way to form Cooper pairs because
the conservation of spin angular momentum would be violated in any normal
conducting state. Apart from this participation of the conservation of spin 
angular momentum, the mechanism of Cooper pair formation within the NHM is
identical to the familiar mechanism presented within the
Bardeen-Cooper-Schrieffer (BCS) theory.\cite{bcs} 

The participation of the conservation law of spin angular momentum may be
interpreted in terms of quantum mechanical constraining forces which
constrain the electrons to form Cooper pairs. There is evidence that these
constraining forces are necessary for the Hamiltonian to have {\em
eigenstates} in which the electrons form Cooper pairs. 

If this is true, then the BCS theory of superconductivity is only
applicable to superconducting bands as defined in the appendix~\ref{app:a},
case (b). When it is applied to other bands, it does not yield the
{\em absolute} energy minimum in the Hilbert space.

\begin{acknowledgments}
  I am indebted to Ernst Helmut Brandt for stimulating discussions on the new
  model.
\end{acknowledgments}

\renewcommand{\theequation}{\Alph{section}\arabic{equation}}

\appendix

\section{Symmetry-adapted Wannier functions}
\label{app:a}
Consider a metal with the space group $G$ and the point group $G_{0}$.  The
elements
\begin{equation}
a = \{\alpha |\vec t\}
\label{a}
\end{equation}
of $G$ consist of a point group operation $\alpha$  and a translation vector
\begin{equation}
\vec t = \vec \tau (\alpha ) + \vec R
\label{t}
\end{equation}
which is the sum of the nonprimitive translation $\vec \tau (\alpha )$
associated with $\alpha$ and a translation vector $\vec R$ of the Bravais
lattice.

The operators $P(a)$ act on a wave function $f(\vec r,t)$  depending on the
position $\vec r$ and the spin coordinate $t$ according to
\begin{equation}
P(a)f(\vec r, t) = f(\alpha^{-1}\vec r - \alpha^{-1}\vec t, \alpha^{-1}t ),
\label{pa}
\end{equation}
where the symbol $\alpha^{-1}t$ is defined in Eq.~\gl{eq:17}.

The effect of $K$ is given by the
equations~\cite{bc}
\begin{equation}
Kf(\vec r) = f^*(\vec r),
\label{k}
\end{equation}
where $f(\vec r )$ stands for any function of position, and
\begin{equation}
Ku_s(t) = g_su_{-s}(t),
\label{kpauli}
\end{equation}
with~\cite{streitwolf}
\begin{equation}
g_{\pm 1/2} = \mp i.
\label{gspauli}
\end{equation}

\subsection*{Case (a): Wannier functions symmetry-adapted to a magnetic
  group} 

Consider a set of $\mu$ energy bands in the paramagnetic band structure of a
metal with $\mu$ atoms per unit cell. The positions of the atoms are still
written as
$$
\vec T = \vec R + \vec\varrho_{i},
$$
where $\vec R$ and $\vec\varrho_{i}$ ($i = 1$ to $\mu$) denote the vectors
of the Bravais lattice and the positions of the $i$th atom within the unit
cell, respectively.

Further, consider the magnetic group 
\begin{equation}
  \label{mg}
M = H + K\{\gamma |\vec\tau (\gamma )\}H  
\end{equation}
where $H$ is a subgroup of $G$,
$$ H \subset G,$$
$K$ denotes the operator of time inversion,
and $\{\gamma |\vec\tau (\gamma )\}$ is a space group element of $G - H$.

Assume degeneracies to exist between the bands of the considered set of $\mu$
energy bands and the bands not belonging to this set. This assumption is
always true since, in any metal, there are degeneracies between the bands of
any selected set of energy bands and the bands outside of this set.
These degeneracies are caused by symmetry and may occur at points and lines
of symmetry of the Brillouin zone. Further, assume
\begin{itemize}
\item these degeneracies to be removed in the subgroup $H$ of $G$
  (i.e., when the representations of $G$ are replaced by the subduced
  representations of $H$);
\item the symmetry operation $K\{\gamma |\vec\tau (\gamma )\}$ not to
  produce extra degeneracies between the bands of the considered set and
  bands outside of this set;
\item unitary matrices ${\bm S}(\vec K)$ [as defined in Eq.~(4.16) of
  Ref.~\onlinecite{ew1}] to exist which satisfy the equations (4.17) and
  (4.28) of Ref.~\onlinecite{ew1} and Eq.~(7.1) of Ref.~\onlinecite{ew3}; and
\item the positions $\vec\rho_i$ of the Wannier functions [which are
  determined by these matrices ${\bm S}(\vec K)$] to be identical with the
  positions of the atoms.
\end{itemize}

Then the coefficients $g_{iq}(\vec k )$ in Eq.~\gl{wf} may be chosen in such
a way that the Wannier functions comply with the four conditions following
Eq.~\gl{wf} with the exception that they are no longer symmetry-adapted to
the paramagnetic space group $G$ but only to the subgroup $H$ of $G$. That
means, Eq.~\gl{swf} is satisfied only for the elements $\alpha$ of the point
group $H_0$ of $H$.
In addition to Eq.~\gl{swf} we have\cite{ew3}
\begin{equation}
  \label{kswf}
Kw_i\big(\gamma^{-1}(\vec r - \vec R - \vec\rho_i)\big) = \sum_{j =
  1}^{\mu}D_{ji}(K\gamma )w_j(\vec r - \vec R - \vec\rho_i) 
\end{equation}
where the matrix $[D_{ji}(K\gamma )]$ is the representative of $K\gamma$ in
the corepresentation of the point group
\begin{equation}
\label{pmg0}
M_0 = H_0 + K\gamma H_0  
\end{equation}
of $M$ which is derived from the representation $D_0$ of $H_0$ in
Eq.~\gl{swf}.  
 
Since there is exactly one Wannier function at each atom, the Wannier
functions may be labeled by the positions $\vec T$ of the atoms,
\begin{equation}
\label{abk}
w_{\vec T}(\vec r) \equiv w_i(\vec r - \vec R - \vec\rho_i),  
\end{equation}
and the equations~\gl{swf} and~\gl{kswf} may be considerably simplified.

Applying on both sides of Eq.~\gl{swf} the operation $\alpha$ on $\vec
r - \vec R - \vec\rho_i$ we obtain
\begin{equation}
  \label{alphaswf}
w_i(\vec r - \vec R - \vec\rho_i) = \sum_{j =
  1}^{\mu}D_{ji}(\alpha )w_j\big(\alpha(\vec r - \vec R - \vec\rho_i)\big), 
\end{equation}
and the application of the operator $P(a)$ [given in Eq.~\gl{pa}] on both
sides of this equation~\gl{alphaswf} yields the equation
\begin{equation}
  \label{pswf1}
P(a)w_i(\vec r - \vec R - \vec\rho_i) = 
\sum_{j = 1}^{\mu}D_{ji}(\alpha )w_j(\vec r - \vec t - \alpha\vec R
-\alpha\vec\rho_i) 
\end{equation}
which applies to all the elements $a \in H$.

As shown in Ref.~\onlinecite{ew2}, Eq.~\gl{pswf1} may also be written in the
form
\begin{equation}
  \label{pswf2}
P(a)w_i(\vec r - \vec R - \vec\rho_i) = 
\sum_{j = 1}^{\mu}D_{ji}(\alpha )w_j\big(\vec r - \alpha\vec R - \vec\rho_j -
\vec R_j(\alpha)\big) 
\end{equation}
with $\vec R_j(\alpha)$ being translations of the Bravais lattice (depending
on $j$ and $\alpha$), see Eq.~(2.13) of Ref.~\onlinecite{ew2}.

Comparing the Eqs.~\gl{pswf1} and \gl{pswf2}, we see that on the right hand
side of Eq.~\gl{pswf2} there are only Wannier functions related to an atomic
position $\vec\rho_j$ within the unit cell for which the translation 
\begin{equation}
  \label{rj}
\vec R_j(\alpha ) = \alpha\vec\rho_i + \vec t - \vec\rho_j   
\end{equation}
is a translation vector of the Bravais lattice. This cannot be true for more
than one vector $\vec\rho_j$ since all the $\vec\rho_j$ are different and lie
within the unit cell. It is true for exactly one $\vec\rho_j$ because $\vec
R_j(\alpha )$ is a translation of the Bravais lattice if we put
\begin{equation}
  \label{rhoj}
\vec\rho_j = \alpha\vec\rho_i + \vec t
\end{equation}
and $\vec R + \alpha\vec\rho_i + \vec t$ is the position of an atom
since it can be generated by the application of the space group operation
$\{\alpha|\vec t\}$ on the atomic position $\alpha^{-1}\vec R + \vec\rho_i$.

Consequently, on the right hand side of Eq.~\gl{pswf1} there is only one
Wannier function, namely that function related to the atom at position
$\alpha\vec\rho_i + \vec t$ within the unit cell. 
Hence, the sum on the right hand side of Eq.~\gl{pswf1} and (analogously)
of Eq.~\gl{kswf} consists of one summand only. The matrices
$[D_{ji}(\alpha )]$ in these equations have only one non-vanishing element,
say $d_{ji}(\alpha )$, in each column which satisfies the equation
\begin{equation}
\label{dji}
|d_{ji}(\alpha )| = 1  
\end{equation}
since the matrix $[D_{ji}(\alpha )]$ is unitary.

Hence, Eq.~\gl{pswf1} may be written as 
\begin{equation}
  \label{pswf3}
P(a)w_{\vec T}(\vec r) = d_{\vec T}(\alpha )w_{\vec T'}(\vec r)\qquad
\mbox{ for } a \in H
\end{equation}
with
\begin{equation}
  \label{t'}
\vec T' = \alpha\vec T + \vec t  
\end{equation}
and Eq.~\gl{kswf} yields
\begin{equation}
  \label{kswf2}
KP(g)w_{\vec T}(\vec r) = d_{\vec T}(K\gamma )w_{\vec T'}(\vec r)
\end{equation}
with
$$g = \{\gamma|\vec\tau (\gamma )\}$$
and
$$\vec T' = \gamma\vec T + \vec\tau (\gamma ),  $$
where the coefficients $d_{\vec T}(\alpha )$ and $d_{\vec T}(K\gamma )$
have the absolute value 1,
\begin{equation}
  \label{dt}
|d_{\vec T}(\alpha )| = |d_{\vec T}(K\gamma )| = 1.
\end{equation}

It should be noted that the time inversion $K$ does not belong to $M$.
Therefore, it is not possible to choose the coefficients $g_{iq}(\vec k )$ in
Eq.~\gl{wf} in such a way that the Wannier functions satisfy an equation
analogous to Eq.~\gl{kswf2} by application of the time inversion operator $K$
{\em alone}.

\subsection*{Case (b): Spin dependent Wannier functions symmetry-adapted to
  the paramagnetic group}  

\subsubsection{Symmetry operators}

If in Eq.~\gl{wf} we replace the Bloch functions $\varphi_{\vec kq}(\vec
r)$ by Bloch functions
\begin{equation}
\phi_{\vec kqm}(\vec r,t) = \sum_{s = -\frac{1}{2}}^{+\frac{1}{2}}
f_{sm}(q,\vec k)u_{s}(t)\varphi_{\vec kq}(\vec r)
\label{sdbf}
\end{equation}
with $\vec k$ dependent spin directions, we get ``spin dependent Wannier
functions''
\begin{eqnarray}
\label{sdwf}
\lefteqn{w_{im}(\vec r - \vec R - \vec\rho_{i}, t)}\nonumber\\*
&=& 
\frac{1}{\sqrt{N}}\sum^{BZ}_{\vec k}
\sum_{q = 1}^{\mu}
e^{-i\vec k (\vec R + \vec\rho_{i})} 
g_{iq}(\vec k)\phi_{\vec kqm}(\vec r,t) ,\nonumber\\
\end{eqnarray}
which are labeled by the additional quantum number $m = \pm\frac{1}{2}$ of
the crystal spin. The functions $u_{s}(t)$ denote Pauli's spin
functions as given in Eq.~\gl{paulisf} 
and the coefficients $f_{sm}(q,\vec k)$ form (for each $\vec
k$ and $q$) a unitary two-dimensional matrix ${\bm f}(q, \vec k)$,
\begin{equation}
  \label{fnk}
{\bm f}^{-1}(q, \vec k) = {\bm f}^{\dagger}(q, \vec k).  
\end{equation}

If we have 
\begin{equation}
  \label{eq:14}
f_{sm}(q,\vec k) = \delta_{sm},  
\end{equation}
the two functions $\phi_{\vec kqm}(\vec r,t)$ (with $m = \pm\frac{1}{2}$) are
usual Bloch functions with the spins lying in $+z$ and $-z$ direction,
respectively. Otherwise, the functions $\phi_{\vec kqm}(\vec r,t)$ still are
usual Bloch functions with antiparallel spins which, however, no longer lie in
$\pm z$ direction.

As in the preceding case (a), consider a set of $\mu$ energy bands
in the paramagnetic band structure of a metal with $\mu$ atoms per unit cell.

The paramagnetic group $M^P$ of the metal may be written as
$$
M^P = G + KG  
$$
where $K$ still denotes the operator of time inversion. 
Assume
\begin{itemize}
\item the symmetry degeneracies between the bands belonging the considered
  set and bands not belonging to this set to be removed when the
  single-valued representations of $G$ are replaced by the corresponding {\em
    double-valued} representations;
\item the time inversion symmetry not to produce extra degeneracies between
  the bands of the considered set and bands outside of this set;
\item unitary matrices ${\bm S}(\vec K)$ to exist which satisfy the equations
  (4.16), (4.17), and (4.28) of Ref.~\onlinecite{ew1} and Eq.~(7.1) of
  Ref.~\onlinecite{ew3}, when the single-valued representations in these
  equations are replaced by the corresponding double-valued representations;
  and
\item the positions $\vec\rho_i$ of the Wannier functions to be identical
  with the positions of the atoms.
\end{itemize}

Then the coefficients $f_{sm}(q,\vec k)$ and $g_{iq}(\vec k)$ in
Eqs.~\gl{sdbf} and \gl{sdwf} may be chosen in such a way that also the spin
dependent Wannier functions comply with the four conditions following
Eq.~\gl{wf}. However, the coefficients $f_{sm}(q,\vec k)$ cannot be chosen
independent of $\vec k$ since the considered set of energy bands is isolated
not before the single-valued representations of the Bloch functions are
replaced by the related double-valued representations. [If the
$f_{sm}(q,\vec k)$ are independent of $\vec k$, the Wannier functions in
Eq.~\gl{sdwf} are usual Wannier functions which may comply with the four
conditions following Eq.~\gl{wf} only if the considered set of energy bands
is already isolated when the Bloch functions are labeled by the single-valued
representations of $G$.] As an important consequence, the operator $H'$ does
not conserve the crystal spin, see Eq.~\gl{eq:12}.

The symmetry of the spin dependent Wannier functions may be derived from the
equations in Refs.~\onlinecite{ew1}, \onlinecite{ew2} and \onlinecite{ew3} in
the same way as we have derived the symmetry of the magnetic Wannier
functions in the preceding case (a).  We now get the equations
\begin{equation}
  \label{ssdwf}
P(a)w_{\vec Tm}(\vec r,t) = d_{\vec T}(\alpha )
\sum_{m' = -\frac{1}{2}}^{+\frac{1}{2}}
d_{m'm}(\alpha )
w_{\vec T'm'}(\vec r,t)
\end{equation}
for $a \in G$, 
and
\begin{equation}
  \label{ksdwf}
Kw_{\vec Tm}(\vec r,t) = d_{\vec T}(K)\sum_{m' = -\frac{1}{2}}^{+\frac{1}{2}}
d_{m'm}(K)
w_{\vec T m'}(\vec r, t),
\end{equation}
where 
$$
w_{\vec Tm}(\vec r, t) \equiv w_{im}(\vec r - \vec R - \vec\rho_i, t)  
$$
and
$$\vec T' = \alpha\vec T + \vec t.$$
The operators $P(a)$ now
act on $\vec r$ and $t$, see Eq.~\gl{pa}, the matrices
$[d_{m'm}(\alpha )]$ are the representatives of the two-dimen\-si\-onal
double-valued representation $D_{1/2}$ of the three-dimen\-si\-onal rotation
group $O(3)$, the matrix $[d_{m'm}(K)]$ is given by
\begin{equation}
  \label{dk}
[d_{m'm}(K)] = \left(
\begin{array}{rr}
0&1\\
-1&0
\end{array}
\right),   
\end{equation}
[see, e.g., Table 7.15 of Ref.~\onlinecite{bc}], and the c-numbers $d_{\vec
  T}(\alpha )$ and $d_{\vec T}(K)$ still have the absolute value 1,
\begin{equation}
  \label{dt2}
|d_{\vec T}(\alpha )| = |d_{\vec T}(K)| = 1.
\end{equation}

The equations~\gl{ssdwf} and \gl{ksdwf} are already given in
Ref.~\onlinecite{em}. Further, in appendix B of Ref.~\onlinecite{em} simple
equations are given to identify sets of energy bands complying with all the
condition given above.

\subsubsection{Operators of the crystal spin}

Define the ``group of the positions $\vec\rho_i$'' $G_M$ to consist of all
the $\alpha \in G_0$ which satisfy the equation
\begin{equation}
  \label{eq:20}
  \alpha\vec\rho_i + \vec\tau(\alpha ) = \vec\rho_i + \vec R_i 
\end{equation}
for each $\vec\rho_i$, where $\vec R_i$ denotes a translation vector of the
Bravais lattice, and define for all $\alpha \in G_M$ symmetry operators of
the ``crystal spin''
\begin{equation}
  \label{eq:19}
  M(\alpha ) = P(\{E|\vec R - \vec R_i\})P[\{\alpha|\tau(\alpha
  )\}]P(\{E|-\vec R\}) 
\end{equation}
which depend on the position
$$
\vec T = \vec R + \vec\rho_i
$$
of the (spin dependent) Wannier function on which they are acting. 

From Eqs.~\gl{eq:19} and \gl{ssdwf} we obtain the equation
\begin{equation}
  \label{eq:21}
M(\alpha)w_{\vec Tm}(\vec r,t) = d_{\vec T}(\alpha )\sum_{m' =
  -\frac{1}{2}}^{+\frac{1}{2}} 
d_{m'm}(\alpha )
w_{\vec Tm'}(\vec r,t)
\end{equation}
for $\alpha \in G_M$, showing that the operators $M(\alpha)$ leave unchanged
the positions of the spin dependent Wannier functions. 

\section{Symmetry of the fermion operators}
\label{app:b}
\subsection{Adiabatic fermion operators}
\label{adiabat}
The adiabatic fermion operators $c_{\vec Tm}^{\dagger}$ and $c_{\vec Tm}$
create and annihilate electrons in localized states represented by the the
Wannier functions $w_{\vec T}(\vec r)u_m(t)$ [in the case (a) of magnetic
Wannier functions] or $w_{\vec Tm}(\vec r,t)$ [in the case (b) of spin
dependent Wannier functions]. Hence, their symmetry is determined by the
equations \gl{eq:17}, \gl{pswf3}, \gl{kswf2}, \gl{ssdwf}, and
\gl{ksdwf}. From these equations we
get
\begin{equation}
P(a)c^{(n)\dagger}_{\vec Tm}P^{-1}(a) = d_{\vec T}(\alpha )\sum_{m' = -\frac{1}{2}}
^{+\frac{1}{2}}
d_{m'm}(\alpha )
c^{(n)\dagger}_{\vec T'm'}
\label{eq:0}
\end{equation}
with
$$
\vec T' = \alpha\vec T + \vec t  
$$
and
\begin{equation}
KP(g)c^{(n)\dagger}_{\vec Tm}[KP(g)]^{-1} = d_{\vec T}(K\gamma )\sum_{m' =
  -\frac{1}{2}} 
^{+\frac{1}{2}} 
d_{m'm}(K\gamma )
c^{(n)\dagger}_{\vec T'm'}
\label{eq:1}
\end{equation}
with
$$
\vec T' = \gamma\vec T + \vec\tau (\gamma ),
$$
where the superscript $(n)$ of the fermion operators should be disregarded in
this section. 
In addition, the operators of the crystal spin satisfy the equation 
\begin{equation}
  \label{eq:22}
M(\alpha)c^{(n)\dagger}_{\vec Tm}M^{-1}(\alpha) = d_{\vec T}(\alpha )\sum_{m' =
  -\frac{1}{2}}^{+\frac{1}{2}} 
d_{m'm}(\alpha )
c^{(n)\dagger}_{\vec Tm'}
\end{equation}
for all the elements $\alpha$ of the group of the positions $G_M$, see
Eqs.~\gl{eq:20} and \gl{eq:19}.

The coefficients $d_{\vec T}(\alpha )$ and $d_{\vec T}(K\gamma
)$ still have the absolute value 1,
$$
|d_{\vec T}(\alpha )| = |d_{\vec T}(K\gamma )| = 1,
$$
and the matrices $[d_{m'm}(\alpha )]$ and $[d_{m'm}(K\gamma )]$ still are
representatives of the two-dimen\-si\-onal double-valued representation
$D_{1/2}$ of the three-dimen\-si\-onal rotation group $O(3)$ and the
corepresentation of $O(3) + KO(3)$ derived from $D_{1/2}$, respectively.

In the case (a) of the magnetic Wannier functions, Eq.~\gl{eq:0} is valid for all
the elements $a$ in $H$,
\begin{equation}
  \label{eq:2}
a \in H,  
\end{equation}
and in Eq.~\gl{eq:1} we have
\begin{equation}
  \label{eq:3}
g = \{\gamma|\vec\tau(\gamma)\}.
\end{equation}
In the case (b) of the spin dependent Wannier functions Eq.~\gl{eq:0} is valid
for all the elements $a$ in $G$,
\begin{equation}
  \label{eq:4}
a \in G,  
\end{equation}
and in Eq.~\gl{eq:1} we have
\begin{equation}
  \label{eq:5}
g = \{E|\vec 0\}  
\end{equation}
where $E$ denotes the identity element of $G_0$. In the latter case the matrix
$[d_{m'm}(K)]$ is given in Eq.~\gl{dk}.

\subsection{Nonadiabatic fermion operators}
\label{nonadiabat}
Within the NHM, the Wannier functions $w_{\vec T}(\vec r)u_m(t)$ [in the case
(a) of magnetic Wannier functions] or $w_{\vec Tm}(\vec r,t)$ [in the case
(b) of spin dependent Wannier functions] are replaced by nonadiabatic
localized functions,
\begin{equation}
  \label{eq:-1}
\left.\begin{array}{l}
w_{\vec T}(\vec r)u_m(t)\\  
w_{\vec Tm}(\vec r,t)
\end{array}
\right\} \longrightarrow
\langle\vec r,t, \vec q~|\vec T,m, n \rangle,
\end{equation}
having the same symmetry as the Wannier functions.
However, in the case of the nonadiabatic localized states, the symmetry
operators $P(a)$ act on $\vec r, t$, and on the new coordinate $\vec q$
according to
\begin{eqnarray}
\lefteqn{
P(a)\langle\vec r,t, \vec q~|\vec T,m, n \rangle}\nonumber\\
&& = \langle\alpha^{-1}\vec r
- \alpha^{-1}\vec t, \alpha^{-1}t, \alpha^{-1}\vec q~|\vec T,m, n \rangle, 
\label{effectp}
\end{eqnarray}
where the symbol $\alpha^{-1}t$ is defined in Eq.~\gl{eq:17},
and the application of $K$ yields
\begin{equation}
K\langle\vec r,t, \vec q~|\vec T,m, n \rangle = g_m\langle\vec r, t, \vec
q~|\vec T,-m, n \rangle^*,
\label{effectk}
\end{equation}
with $g_m$ being
given in Eq.~\gl{gspauli}.

With these redefinitions of the symmetry operators, the symmetry of the
nonadiabatic fermion operators $c^{n\dagger}_{\vec Tm}$ is also given by the
equations of the preceding appendix~\ref{adiabat}. The superscript $(n)$ of the
fermion operators in these equations shall indicate that they are valid for
both the adiabatic and nonadiabatic fermion operators.


\begin{thebibliography}{10}
\bibitem{bem2}
\bibinfo{note}{With free atoms, the nonadiabatic motion of the nuclei occurs in
  such a way that the total centre of mass stays at rest. In a solid state,
  consisting of bound atoms, this restriction does not hold.}
\bibitem{bem3}
\bibinfo{note}{The statement ``$E(\alpha)$ is not true for at least one $\alpha
  \in G_0$'' is the negation of ``$E(\alpha)$ is true for $\alpha \in G_0$'',
  where $E(\alpha)$ stands for any equation depending on $\alpha$. It is
  meaningless for how many elements of $G_0$ the equation $E(\alpha)$ is not
  true.}
\end{thebibliography}

\end{document}